# Ultrafast optical control of polariton energy in an organic semiconductor microcavity

Kirsty E. McGhee[1], Michele Guizzardi[2], Rahul Jayaprakash[1], Kyriacos Georgiou[1,3], Till Jessewitsch[4], Ullrich Scherf[4], Giulio Cerullo[2,5], Anton Zasedatelev[6], Tersilla Virgili[5], Pavlos G. Lagoudakis[6] and David G. Lidzey[1]*

1. Department of Physics and Astronomy, University of Sheffield, Hicks Building, Hounsfield Road, Sheffield, S3 7RH, UK
2. Dipartimento di Fisica, Politecnico di Milano, Piazza Leonardo da Vinci 32, 20133 Milano, Italy
3. Department of Physics, Laboratory of Ultrafast Science, University of Cyprus, Nicosia 1678, Cyprus
4. Macromolecular Chemistry Group and Wuppertal Center for Smart Materials & Systems (CM@S), Bergische Universität Wuppertal, Gauss-Strasse 20, 42119 Wuppertal, Germany
5. Istituto di Fotonica e Nanotecnologia-CNR, Piazza Leonardo Da Vinci 32, 20133 Milano, Italy
6. Department of Physics and Astronomy, University of Southampton, University Road, Southampton SO17 1BJ, UK

* Corresponding author: d.g.lidzey@sheffield.ac.uk

**Abstract** The manipulation of exciton-polaritons and their condensates is of great interest due to their applications in polariton simulators and high-speed, all-optical logic devices. Until now, methods of trapping and manipulating such condensates are not dynamically reconfigurable or result in an undesirable reduction in the exciton-photon coupling strength. Here, we present a new strategy for the ultrafast control of polariton resonances via transient modification of an optical cavity mode. We have constructed multilayer organic semiconductor microcavities that contain two absorbers: one strongly-coupled to the cavity photon mode and one that is out-of-resonance. By selectively exciting the out-of-resonance absorber using ultrashort laser pulses, we modulate the cavity refractive index and generate fully-reversible blueshifts of the lower polariton branch by up to 8 meV in sub-ps timescales with no corresponding reduction in the exciton-photon coupling strength. Our work demonstrates the ability to manipulate polariton energy landscapes over ultrafast timescales with important applications in emerging computing technologies.

## 1. Introduction

Exciton-polaritons are bosonic light-matter quasiparticles consisting of an exciton – a bound electron-hole pair – coupled to a microcavity-photon mode. They can be evidenced by the appearance of two branches in the cavity reflectivity termed the lower (LPB) and upper (UPB) polariton branches that undergo an anti-crossing at the point of exciton-photon resonance. The energy-momentum dispersion relation at the bottom of the LPB (around $k = 0$) has a near parabolic dependence, with polaritons able to relax in energy and momentum to this point following thermalisation. If a sufficient density of polaritons collect at the bottom of the LPB branch, the overlap of their wavefunctions can result in an out-of-equilibrium Bose-Einstein-like condensation process. In this macroscopic state – known as a polariton condensate – all the polaritons have the same energy, momentum and phase. These condensates have been widely studied in microcavities containing both inorganic (III-V,[1–5] II-VI[6,7]) and organic semiconductors, with polariton condensates observed at room temperature in organic-based systems due to the high binding energy of Frenkel excitons.[8–13]

In recent years, there has been a significant focus on trapping polariton condensates through the spatial confinement of their excitonic or photonic components. Such traps alter the energy and density of the polaritons, allowing control over their transport, interactions and scattering mechanisms. As well as providing a basis for fundamental studies, this effect has potential applications in nonlinear photonic integrated circuits and polariton logic devices.[14–19] Due to the low effective mass of polaritons ($10^{-4}$ to $10^{-5}$ times that of an electron[20,21]) and their large propagation distances, devices based on polariton condensates are expected to be extremely fast and efficient, with reported polariton velocities of up to 150 μm ps$^{-1}$ and sub-femtojoule switching energies.[17,22,23] The use of multiple traps arranged in lattices has also been explored due to their ability to simulate complex many-body phenomena.[24–26] A range of physical techniques can be used to confine the photonic component of a polariton; for example, using micropillars fabricated through lithography and etching,[27–29] or via Gaussian defects fabricated through laser patterning.[30] Recently, a series of focused laser beams have been used to drive localised regions in an organic semiconductor microcavity into the weak coupling regime forming a localised barrier region, thereby enabling the trapping of condensates in strongly-coupled regions that were located close to the barrier.[31] The excitonic component of a polariton can also be trapped through the application of mechanical strain,[21,32,33] or by using structured laser beams.[34,35] The advantage of optical control techniques is that such traps can be dynamically modified, allowing an additional degree of control over polariton condensate dynamics.

In inorganic semiconductor microcavities, the optical generation of traps generally relies on the strong Coulombic repulsion between polaritons and the uncoupled excitons (the so-called exciton reservoir), which creates an energetic blueshift. In organic semiconductor cavities, however, the blueshift of the LPB emission observed at the condensation threshold arises from two mechanisms that occur when the oscillator strength of the organic semiconductor is reduced.[36] The first mechanism results from a reduction of the Rabi splitting energy (the minimum energetic separation between the UPB and LPB) and is proportional to the square root of the oscillator strength of the excitonic transition.[37] Here, the transient bleaching of the organic material absorption results in a reduced Rabi splitting causing an energetic blueshift of the LPB. The second mechanism results from an effective change in cavity refractive index that occurs as a result of pumping an electronic transition. Here, by depleting the ground state using an intense optical pump, it is possible to reduce the effective refractive index of a material at wavelengths longer than that of the electronic transition. This effect leads to a reduction in the effective optical path length of the cavity with an associated blueshift of the LPB (which is located at lower energy with respect to the strongly-coupled excitonic absorption transition).

In this work, we have fabricated organic semiconductor microcavities that contain separate, non-interacting strongly-coupled and out-of-resonance molecular materials. The out-of-resonance material, binaphthyl-polyfluorene (BN-PFO), was chosen to have an excitonic transition whose energy is located in energy above that of the strongly-coupled material. The out-of-resonance material was then pumped using ultrashort laser pulses to partially bleach its ground state; an effect that reduced the cavity optical path length and resulted in a fully-reversible energetic blueshift of the LPB of up to 8 meV on sub-ps timescales. We show using transfer matrix reflectivity (TMR) modelling that an excited state absorption evident in the BN-PFO around the spectral region of the LPB additionally increases the blueshift through a concomitant reduction in refractive index. Our approach is rather different from that outlined by Wei *et al* in ref [31], in which optical pumping of a cavity was used to create a blueshifted barrier region having reduced exciton-photon coupling strength. Such barrier regions were able to confine polariton condensates that were trapped in strongly-coupled regions adjacent to the barriers. In our approach, we create blueshifted barrier regions in which there is no reduction in exciton-photon coupling strength, with all parts of the cavity (including the barrier regions) remaining in the strong coupling regime. We believe our approach has potential advantages, as it presents a method by which polariton condensates could be made to move between sites of different potential in an optically-printed lattice. Indeed, we show that the cavity energy landscape can be modified over ultrafast timescales and can be partially recovered in a few picoseconds, potentially allowing the creation of new types of polaritonic logic devices and ultrafast optical switches. We emphasise, however, that in contrast to work described in ref [31], we do not generate polariton condensates, but instead focus on the photophysics of the confining potential that could be used to trap a condensate.

## 2. Results and discussion
### 2.1 Molecular materials and multilayers

Microcavities were fabricated containing a layer of the strongly-coupled molecular dye BODIPY-Br (see chemical structure in Figure 1(a)). This material (and its derivatives) have been widely studied in organic microcavities, with such cavities being able to undergo polariton condensation[11,38,39]. BODIPY-Br was dispersed in a polystyrene (PS) matrix to reduce intermolecular interactions that result in linewidth broadening and quenching of photoluminescence (PL). The normalised absorption and PL spectra of BODIPY-Br dispersed at 10% by mass in PS are shown in Figure 1(b). The optical absorption peaks at 530 nm and has a linewidth of 24 nm (106 meV) (full width at half maximum (FWHM)). The PL (excited using a 405 nm CW laser diode) has a peak at 548 nm with a shoulder around 592 nm which has been ascribed to a combination of emission from a vibronic-replica and molecular aggregates.[40,41]

The conjugated polymer BN-PFO was placed into the same microcavity, but was out-of-resonance with the cavity mode as a result of the large energetic detuning between its excitonic peak and the cavity mode resonance. The BN-PFO copolymer consisted of a random mixture of 2,7-(9,9-di-*n*-octylfluorene) and a small fraction (9.8 Mol%) of 6,6'-(2,2'-*n*-octyloxy-1,1'-binaphthyl) repeat units (see chemical structure in Figure 1(a)). The absorption and PL (again generated following CW excitation at 405 nm) spectra of BN-PFO are shown in Figure 1(c). It can be seen that its absorption is relatively broad (FWHM of 68 nm) and peaks at 375 nm. The PL spectrum is characterised by an electronic transition at 421 nm together with vibronic replicas at 446 and 470 nm. BN-PFO is highly photostable, and has been shown to undergo quasi-continuous-wave lasing.[42,43] Note that the absorption of the BN-PFO is located at a shorter wavelength than that of the BODIPY-Br; however, it is sufficiently close to the LPB that on optical excitation of the BN-PFO, the LPB is expected to experience a change in the local refractive index. Clearly if the spectral separation between the BN-PFO and the LPB were increased significantly, any changes in refractive index experienced by the LPB that were generated by a transient bleach of the BN-PFO would be reduced.

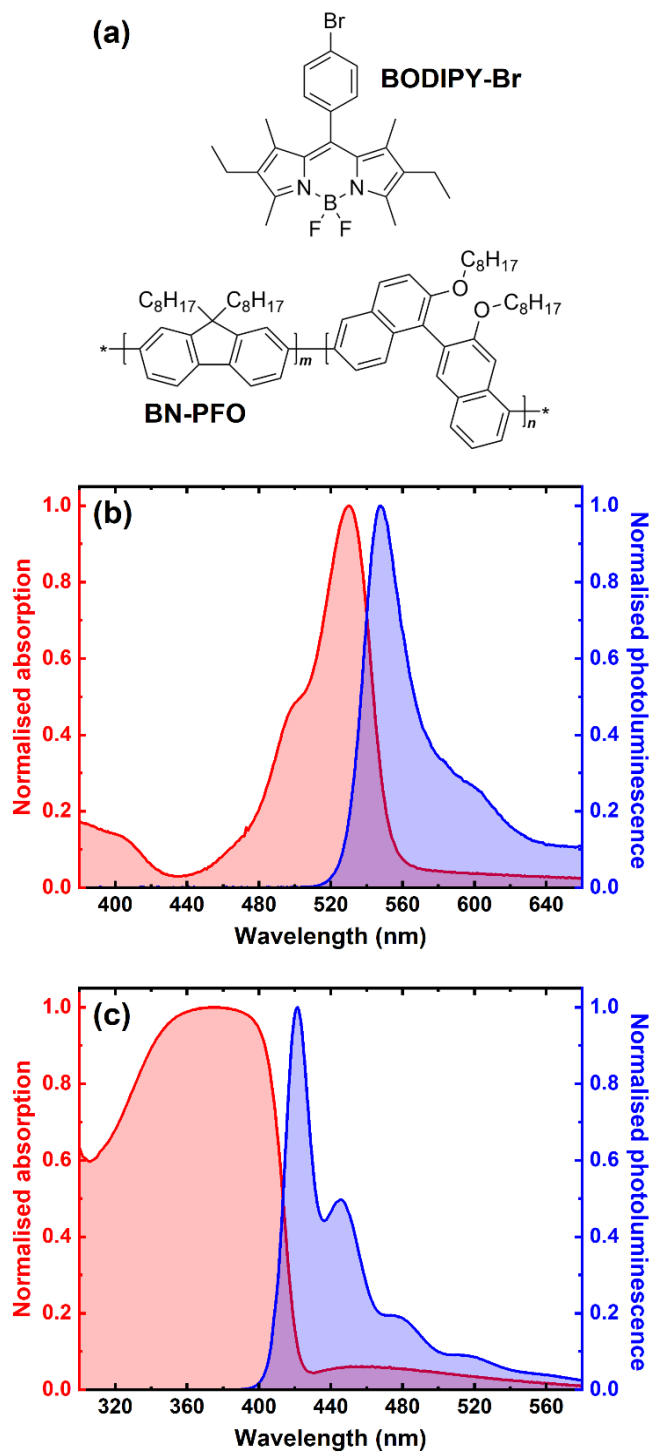

**Figure 1.** Characterisation of the BODIPY-Br and BN-PFO films. Part (a) shows the molecular structure of BODIPY-Br (top) and BN-PFO (bottom). In the BN-PFO used here, $\frac{n}{(n+m)} = 9.8\%$, where $m$ and $n$ represent the relative number of PFO and BN units, respectively. Parts (b) and (c) show the normalised absorption

(red) and photoluminescence (blue) spectra of BODIPY-Br and BN-PFO, respectively. In part (c), scatter from the laser in the PL data was removed post-processing. Note that the weak but relatively broad background evident at longer wavelengths in the BN-PFO absorption results from an optical interference effect, rather than being generated by a direct electronic transition.

When placed into a microcavity, BODIPY-Br and BN-PFO were separated by an inert, 60 nm thick polyvinyl alcohol (PVA) spacer layer. The function of this spacer layer was to prevent direct electronic interaction (such as the formation of charge-transfer states) or dipole-dipole energy transfer between the BN-PFO and BODIPY-Br. We are confident that such interaction processes are excluded here, as the thickness of the PVA barrier layer was much larger than the typical Forster transfer radius between a molecular dye and a PFO-based polymer (around 5 nm [44]).

Such multilayer films could be created using solution processing due to the differing solubilities of the materials. Here, BODIPY-Br (dispersed in PS) and BN-PFO are only soluble in common organic solvents such as toluene, while PVA is only soluble in aqueous (polar) solvents. This allowed PVA to be spin-coated onto a BODIPY-Br film without causing it to dissolve, with BN-PFO then being spin-coated on top of the PVA, forming a stable, well-defined multilayer.

We firstly discuss the optical properties of multilayer films outside of a microcavity. Here, the multilayer had the structure BODIPY-Br / PVA / BN-PFO, with individual layers having thicknesses of 285 nm / 60 nm / 170 nm, respectively (see schematic in Figure 2(a)). These thicknesses were chosen on the basis of a TMR model that suggested that the BODIPY-Br would undergo strong coupling when placed into a microcavity, with the LPB positioned around 570 nm at normal incidence – the wavelength around which BODIPY-Br undergoes polariton condensation.

The absorption and PL spectra of the multilayer are shown in Figure 2(b). It can be seen that its absorption can be described as a superposition of the absorption of BODIPY-Br and BN-PFO. We find that there is significant PL emitted by the BODIPY-Br following excitation at 405 nm, even though its optical density at this wavelength is low. Indeed, control measurements (see Figure S1 in Supplementary Information) indicate that the PL intensity from the BODIPY-Br in the multilayer film is in fact stronger than that emitted from the pure BODIPY-Br film when excited under comparable conditions. We believe that this results from the PL emitted by the BN-PFO being reabsorbed by the BODIPY-Br and then reemitted. This is also confirmed by the relatively reduced intensity of the BN-PFO PL from the multilayer film compared to the pure BN-PFO film.

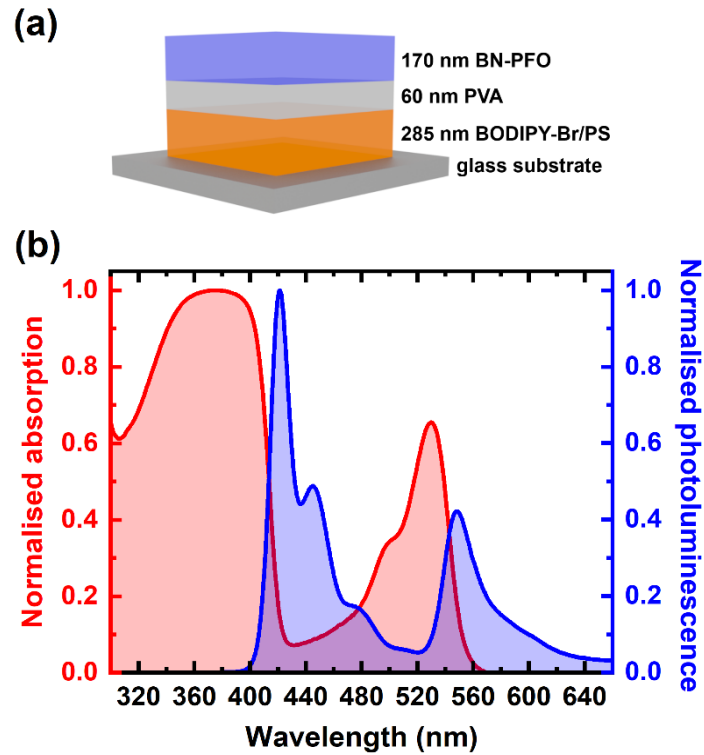

**Figure 2.** Multilayer film. Part (a) shows a schematic of the BODIPY-Br / PVA / BN-PFO film. Part (b) shows its normalised absorption and photoluminescence (PL) (following excitation at 405 nm through the glass substrate) spectra, with PL measured from the back side (i.e. also through the substrate). Both spectra are superpositions of the individual BN-PFO and BODIPY-Br film spectra.

**2.2 Design and fabrication of multilayer microcavities**

We used a TMR model to design a microcavity composed of two distributed Bragg reflectors (DBRs) placed on either side of the multilayer film presented above. A schematic of the cavity structure is shown in Figure 3(a). Here, the bottom DBR consisted of 10 pairs of $SiO_2/Nb_2O_5$ and the top of 8 pairs of $SiO_2/TiO_2$, with both mirror reflectivities centred at 560 nm. Here, by placing 8- and 10-pair DBR mirrors either side of the cavity region, it was possible to create a cavity having one mirror of lower reflectivity through which light could be more easily coupled into and out of the cavity. As input to our model, we have calculated the real ($n$) and imaginary ($k$) part of the complex refractive index of BODIPY-Br by fitting its absorption spectrum to a series of Lorentzian functions with a background refractive index used that approximated the average refractive index of the PS matrix ($n = 1.59$). The $n$ and $k$ values for BN-PFO were instead calculated by applying a Kramers-Kronig relation to its absorption spectrum, with $n =$ 1.8 used as a background refractive index. This value was extracted from fitting a TMR model to the photon dispersion in weakly-coupled cavities containing BN-PFO placed between two semi-transparent silver mirrors.

The TMR model indicated that the BODIPY-Br would strongly-couple to the cavity mode with a Rabi splitting of 103 meV at an external viewing angle of 41°, where the peak of the BODIPY-Br absorption and the cavity photon mode are degenerate in energy. The model also predicted the BN-PFO would remain uncoupled due to its broad linewidth and large positive detuning from the cavity photon mode. Figure 3(b) plots the angle dependent cavity reflectivity predicted by the model could be expected, our TMR model also indicated that any change (reduction) in the extinction coefficient of the BN-PFO layer should result in a reduction in its refractive index at wavelengths > 375 nm (see Figure S3 of the Supplementary Information), thereby causing the cavity mode and hence LPB to blueshift.

To test this concept, we have fabricated the microcavity illustrated in Figure 3(a) (see Methods for a detailed fabrication description). The angle dependent white light reflectivity of the cavity was recorded using a goniometer system and is shown in Figure 3(c). Here, we overlay experimental data (shown using the colour map) with the energy of the polariton modes as predicted by the TMR model (part (b), shown using black crosses). The model provides an excellent fit to the data and indicates a Rabi splitting of 103 meV between the UPB and LPB. If we compare this with the BODIPY-Br exciton linewidth (106 meV FWHM) and the cavity photon linewidth (3.2 meV FWHM), the cavity is comfortably within the strong coupling regime as defined by the criterion introduced by Savona *et al.* [45]. Waterfall plots of the same TMR and experimental reflectivity data are shown in Figure S2 of the Supplementary Information.

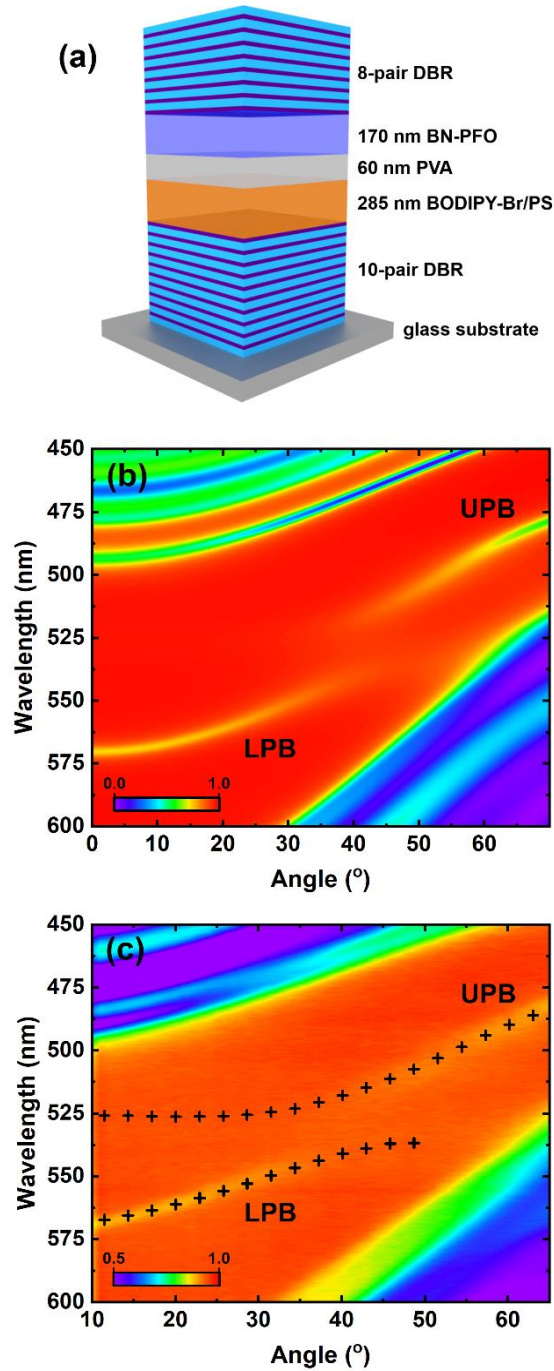

**Figure 3.** BODIPY-Br / PVA / BN-PFO multilayer cavities. Part (a) shows a schematic illustration of the cavity. Part (b) is a TMR simulation of the cavity reflectivity, showing a clear splitting at an angle of 41°. Part (c) shows an experimentally measured cavity reflectivity spectra measured as a function of angle. The results of a TMR model (black crosses) are overlaid on the experimental data. In parts (b) and (c), we identify the upper and lower polariton branches via the labels UPB and LPB, respectively.

## 2.3 Pump probe spectroscopy

To explore whether a measurable blueshift of the LPB can be observed in practice, we have performed ultrafast pump probe spectroscopy on both the microcavity and the multilayer control films. Since the control films had a relatively high transmission at optical frequencies, the transient measurements were performed in a transmission configuration. However, microcavities had a low optical transmission due to their highly reflective DBR mirrors, and thus cavity structures were measured in a reflection configuration. For this reason, all measurements on films are discussed in terms of the fractional change in the *transmitted* probe signal, $\frac{\Delta T}{T}$, while the measurements on microcavities are discussed in terms of the fractional change in the *reflected* probe signal, $\frac{\Delta R}{R}$. We can however generalise and write

$$\frac{\Delta X}{X}(t) = \frac{X_{ON}(t) - X_{OFF}}{X_{OFF}} \tag{1}$$

where $X_{ON}(t)$ and $X_{OFF}$ are the reflected ($R$) or transmitted ($T$) probe signal with and without the optical pump, respectively, and $t$ is the pump-probe delay. In both types of measurement, samples were pumped at 400 nm from the "top side" (i.e. not through the glass substrate) using 100 fs pulses at 1 kHz and probed with time-delayed broadband white light pulses at 2 kHz. Further details can be found in Methods with an example probe spectrum shown in Figure S4 of the Supplementary Information.

**2.3.1 Measuring transient transmission of multilayer films**

In this section, we discuss transient measurements made on multilayers and their individual components. Specifically, we compared a 285 nm BODIPY-Br film, a 170 nm BN-PFO film, and a BODIPY-Br / PVA / BN-PFO multilayer in which the individual layers had a thickness of 285 nm / 60 nm / 170 nm respectively. All three structures were excited at 400 nm with a relatively high pump fluence of 35 $\mu$J/cm$^2$. The $\frac{\Delta T}{T}$ spectrum was recorded as a function of time delay over the window -0.5 ps to 290 ps. Here, a negative delay corresponds to the probe pulse arriving before the pump, while a positive delay corresponds to the probe pulse following the pump.

Figure 4(a) plots the $\frac{\Delta T}{T}$ spectrum as a function of wavelength for all three control films at a delay of 250 fs. This delay corresponds to the point of maximum or near maximum $\frac{\Delta T}{T}$ signal for all films. We firstly consider the transient transmission of BN-PFO (blue line in the figure). Here it can be seen that $\frac{\Delta T}{T}$ is positive over the wavelength range 420 - 510 nm, with a peak observed at 445 nm and a shoulder at 467 nm. These closely coincide with the peak at 446 nm and shoulder at 470 nm observed in the PL spectrum shown in Figure 1(c). We conclude therefore that over this wavelength range, the transient signal is dominated by BN-PFO stimulated emission (SE). At wavelengths longer than 510 nm, $\frac{\Delta T}{T}$ becomes increasingly negative, a result consistent with excited state absorption (ESA) by singlet excitons[46,47] or by interchain charged states.[48] Similar measurements performed by Xu *et al.* on a similar material (PFO) over a wider spectral range reported that this ESA signal increases up to 800 nm followed by a decrease at wavelengths > 850 nm[49].

The BODIPY-Br $\frac{\Delta T}{T}$ spectrum is shown using an orange line. Here, a positive peak is observed at 540 nm that is likely a combination of ground state bleaching (GSB), corresponding to a peak of the absorption spectrum at 530 nm, and SE, corresponding to the peak of the PL spectrum at 548 nm. The relatively small magnitude of the BODIPY-Br $\frac{\Delta T}{T}$ signal compared to that of the BN-PFO is consistent with the much larger absorption coefficient of BN-PFO at the pump wavelength (400 nm). A small negative signal around 440 nm is also evident, which has previously been attributed to ESA by triplet states in similar BODIPY derivatives[50–52].

The $\frac{\Delta T}{T}$ spectrum of the multilayer is plotted using a purple line. As anticipated, this closely matches the BN-PFO transient transmission, although small differences are seen between the spectra at wavelengths longer than 530 nm. The origins of these differences are not fully understood. We note that spectral features resulting from BODIPY GSB and SE are not evident in the multilayer spectrum. This results from the low optical density of BODIPY-Br at 400 nm, with optical excitation at this wavelength only generating a small fraction of *excited* BODIPY-Br molecules.

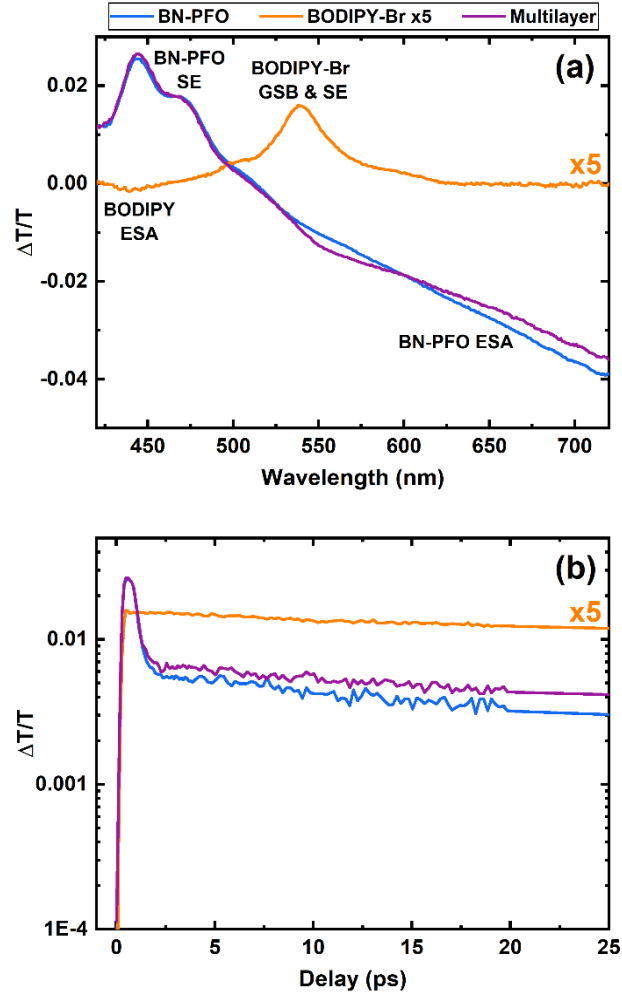

**Figure 4.** Transient absorption measurements on BN-PFO and BODIPY films. Part (a) shows the differential transmission spectra, $\frac{\Delta T}{T}$, for the BN-PFO (blue), BODIPY-Br (orange), and multilayer (purple) films at a pump fluence of 35 $\mu$J/cm$^2$, with the BODIPY-Br spectrum scaled by 5 times to improve clarity. The delay time between pump and probe pulses for all spectra is 250 fs. At this delay, the peaks in the $\frac{\Delta T}{T}$ spectra (445 nm in the BN-PFO and multilayer films and 540 nm in the BODIPY-Br film) were at a maximum or very near their maximum level. Part (b) shows the time dynamics of the $\frac{\Delta T}{T}$ signal at 445 nm in the BN-PFO and multilayer films (corresponding to the peak of the BN-PFO SE) and at 540 nm in the BODIPY-Br film (resulting from a combination of GSB (main absorption peak at 530 nm) and SE (main PL peak at 548 nm)). Again, the BODIPY-Br data has been scaled 5 times for clarity. The labels GSB, SE, and ESA correspond to ground state bleaching, stimulated emission and excited state absorption, respectively.

Figure 4(b) plots the intensity of the $\frac{\Delta T}{T}$ signal as a function of delay time (up to 25 ps) recorded at 445 nm for the BN-PFO and multilayer films and at 540 nm for the BODIPY-Br film. For all films, the $\frac{\Delta T}{T}$ signal rises to its maximum value instantaneously (within our time resolution), after which it decays. For the BN-PFO and multilayer films, we evidence a very fast initial component of the signal decay, having a lifetime of around 290 fs, which we attribute to efficient exciton-exciton annihilation[46,47]. As we describe below, such ultrafast rise and decay dynamics are reflected by rapid changes in the energy of the LPB in a strongly-coupled microcavity.

For completeness, we show equivalent data to that shown in Figure 4 at a lower excitation fluence of 7 µJ/cm$^2$ in Figure S5.1 of the Supplementary Information. Here, we find that the transient spectra for all films are comparable to Figure 4(a) but with a lower amplitude as expected. However, the fast initial decay seen in the BN-PFO and multilayer dynamics (Figure 4(b)) are not evident at this fluence due to the significantly lower exciton density. In Figure S5.2 of the Supplementary Information, we plot the normalised dynamics of all films at both fluences up to a delay time of 290 ps. We find that the $\frac{\Delta T}{T}$ signals of the BN-PFO and multilayer films become negative at longer delay times. This is likely due to the superposition of the BN-PFO SE with an ESA signal having a longer lifetime. We discuss this effect in more detail in the SI.

**2.3.2 Measuring transient reflectivity of microcavities**

Having established the basic photophysics of the multilayer film and component materials, we now discuss transient reflectivity measurements performed on microcavities. Here, our goal was to determine whether we could generate an instantaneous blueshift of the LPB by pumping the BN-PFO at 400 nm. In the data presented here, the reflectivity was probed at an angle of 10º to the cavity normal, with the pump beam nearly collinear to the probe. Measurements were also carried out at higher angles, where the phenomenology was found to be broadly similar.

Figure 5(a) plots the wavelength-dependent transient reflectivity ($\frac{\Delta R}{R}$) of the microcavity containing the multilayer film at a range of pump fluences. Here, we account for the finite transmission of the pump beam through the top DBR, with each given fluence being the calculated fluence of the pump beam inside the microcavity, allowing a direct comparison to data shown in Figures 4, S4 and S5. In Figure 5(a), the data corresponds to the maximum $\frac{\Delta R}{R}$ signal (detected at a delay time of 750 - 800 fs, due to the effect of photon trapping inside the microcavity) around the spectral region of the LPB (568 nm). At each fluence, the transient reflectivity signal takes the form of a derivative lineshape that is negative at wavelengths < 566 nm and positive at wavelengths > 568 nm. This derivative lineshape results from a blueshift of the LPB caused by the optical pump; specifically, there is a reduced reflectivity (negative $\frac{\Delta R}{R}$) at the wavelength to which the LPB has shifted, and increased reflectivity (positive $\frac{\Delta R}{R}$) at the wavelength at which the LPB peaked before the pump.

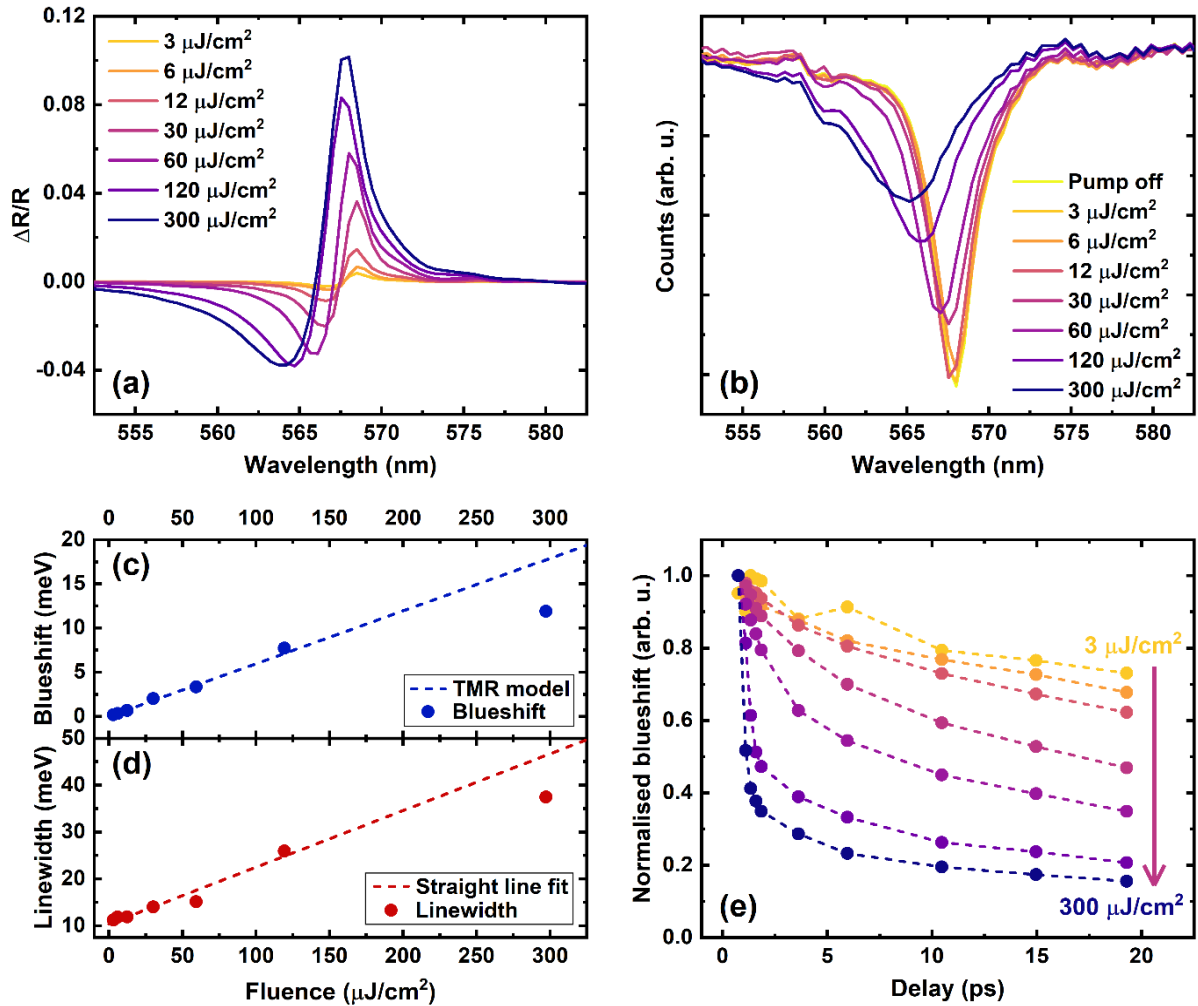

**Figure 5.** Transient reflectivity measurements on BODIPY-Br / PVA / BN-PFO microcavity at an angle of 10° to the cavity normal. Part (a) shows the differential reflectivity, $\frac{\Delta R}{R}$, at different excitation fluences at a pump-probe delay of 750 - 800 fs. The derivative shape is characteristic of a blueshift of the mode – in this case, the LPB. Part (b) shows the reflected probe (with baseline subtraction) at different fluences at a pump-probe delay of 750 - 800 fs, showing that there is indeed a blueshift of the LPB as the excitation fluence is increased. Part (c) shows the measured blueshift as a function of pump fluence (circles) alongside a TMR model (dashed line), showing evidence of degradation at the highest fluence. Part (d) shows the linewidth of the LPB as a function of pump fluence (circles) with a straight line fit to the first six data points (i.e. excluding the highest fluence data point) as a guide to the eye, with degradation again evident at the highest pump fluence. The experimental data in parts (c) and (d) were extracted using a Lorentzian fit to the LPB shown in part (b). Part (e) shows the relaxation of the blueshift towards its unperturbed level as a function of time following the initial optical pump. The legend in part (a) also applies to part (e).

We can in fact directly evidence the blueshift of the LPB from the delay-dependent reflected probe signal $R_{ON}(t) = R_{OFF}(1+\frac{\Delta R}{R}(t))$ recorded at different pump fluences at a pump-probe delay of 750 - 800 fs. This is shown in Figure 5(b). The plot also shows the probe signal recorded in the absence of the pump. (Note, this data was corrected using a baseline subtraction in order to later fit a Lorentzian to the LPB.) From Figure 5(b), it can be seen that when there is no pump present, there is a negative dip in the reflected probe beam at 568 nm corresponding to the LPB. However, when the pump is present, the LPB undergoes a blueshift due to a reduction in the refractive index of the BN-PFO which results in a *reduction in effective cavity path length*. It is apparent that this blueshift is accompanied by a broadening of the LPB.

To quantify the magnitude of the blueshift and linewidth broadening as a function of fluence, a Lorentzian function was fit to the data in Figure 5(b), with the fitted parameters shown in Figures 5(c) and (d), respectively. In Figure 5(c) it can be seen that the magnitude of the LPB blueshift is linearly dependent on the pump fluence for fluences ≤ 120 μJ/cm². At higher fluences, the blueshift increases in a sub-linear fashion, shifting by a maximum of 12 meV (3.1 nm) when excited with an internal cavity fluence of 300 μJ/cm². We have used our TMR model (the dashed line in the figure) to predict the behaviour of the blueshift with increasing fluence, assuming that the reduction in the BN-PFO extinction coefficient is directly proportional to pump fluence. While the model is in excellent agreement with experimental data for low fluences, the observed blueshift (12 meV) is smaller than expected for a pump fluence of 300 μJ/cm². We propose that this effect results from an irreversible photodegradation of the BN-PFO which causes a permanent reduction in its oscillator strength, a conclusion supported by a repeat pump probe scan in which the $\frac{\Delta R}{R}$ signal was found to be reduced. Such degradation was not seen at lower fluences. We conclude therefore that fully reversible blueshifts of up to 8 meV can be generated without noticeable photodegradation.

Figure 5(d) plots the linewidth of the LPB as a function of pump fluence. Here it can be seen that the LPB broadens as the pump fluence is increased (the origin of which we discuss below), going from 11 meV (no pump) to 37 meV (pump fluence of 300 μJ/cm² inside the cavity). Again, this trend is linear up to a fluence of 120 μJ/cm², with a departure from linearity observed at a pump fluence of 300 μJ/cm². This effect is highlighted in the figure using a straight line fit which has been applied to all but the last data point. We again attribute the departure from linearity at a fluence of 300 μJ/cm² to irreversible photodegradation of the BN-PFO.

Figure 5(e) plots the normalised blueshift of the LPB as a function of delay time for the different pump fluences. Such transients were again determined by reconstructing the cavity reflectivity ($R_{ON}(t)$) from the $\frac{\Delta R}{R}$ signal and the probe without the pump ($R_{OFF}$). We note that a similar method to that used here was reported in ref [53]. This procedure allows us to extract the LPB peak wavelength and therefore follow the blueshift as a function of time after the pump pulse. It can be seen that the relaxation of the blueshift becomes progressively faster as the pump fluence increases. Furthermore, we find that the LPB does not fully return to its original position over the time delay range studied. Rather, the blueshift 'switches off' by more than 50% in less than 1 ps at a fluence of 120 µJ/cm$^2$, an effect we attribute to non-linear exciton-exciton annihilation effects in BN-PFO that rapidly modulate its refractive index. For a more quantitative discussion of the decay timescales, see Section 6 of the Supplementary Information. Such ultrafast dynamics are also evident in the blueshift 'switch-on' time, which occurs in around 750 fs. As the rise time of the $\frac{\Delta T}{T}$ signal in the BN-PFO and multilayer films is effectively instantaneous (i.e. at the limit of our experimental resolution), the rise time in the cavity is determined by the cavity quality factor ($Q \sim 2600$).

We now discuss the origin of the observed broadening of the LPB as a function of pump fluence. We propose that this effect results from the excited state absorption transition (ESA) of the BN-PFO. Indeed, Figure 4(a) suggests ESA from BN-PFO around the spectral region corresponding to the LPB is significant. This absorption will clearly act as a loss mechanism in the cavity, resulting in a reduced cavity $Q$-factor and an increased LPB linewidth. We can provide a semi-quantitative description of this effect by including a Lorentzian absorber into our TMR model having a peak at 720 nm and a FWHM linewidth of 600 meV. By reducing the relative BN-PFO oscillator strength by 12% (i.e. to 88% of its unperturbed level) and adjusting the magnitude of the Lorentzian representing the BN-PFO ESA, we are able to reproduce the 8 meV blueshift and 2.1 times broadening of the LPB linewidth as observed in our experiments. Significantly, when the ESA is not included in the model, the blueshift of the LPB is limited to a maximum value of 3 meV at an effective pump fluence of 120 µJ/cm$^2$. This in fact suggests that the majority of the observed blueshift in fact originates from the presence of the ESA transition, which reduces the BN-PFO refractive index around the wavelength of the LPB, rather than from the direct bleaching of the BN-PFO ground state. This simulation is shown and discussed in Section 7 of the Supplementary Information.

It is evident that the optical losses associated with this ESA will likely quench optical gain and could suppress polariton condensation. We do not believe this issue is necessarily problematic, as the effects explored here could be to be exploited to write the 'barriers' that would *confine* a condensate, rather than defining the regions in which condensation actually occurs (i.e. the regions in which the condensated are localised are physically separated from the barrier regions in which there is increased loss).

Finally, we have explored whether the optical losses associated with the ESA are in fact sufficient to drive the cavity into the weak coupling regime. As we show in Section 7 of the Supplementary Information, our TMR simulation of the cavity reflectivity indicates that even at the highest excitation fluence used, the cavity remains strongly-coupled. Indeed, our model indicates that the Rabi splitting energy should increase as a function of increasing excitation fluence. This effect occurs as the reduced optical path length of the photo-pumped cavity will result in an effective increase in the oscillator strength of the BODIPY-Br transition per unit length which thereby increases the Rabi splitting energy (see also analytical expressions for Rabi splitting as a function of optical path length given in ref [54]). It would, however, be interesting to explore whether ESA losses could be reduced by designing a microcavity structure in which the LPB is positioned around 500 nm, a wavelength at which the BN-PFO ESA is significantly reduced (see Figure 4), although such a modification is likely to reduce the blueshift.

### 3. Conclusions

We have fabricated multilayer strongly-coupled organic semiconductor cavities containing the molecular dye BODIPY-Br in which we control the energy (blueshift) of the lower polariton branch. This effect occurs by selectively bleaching an out-of-resonance absorber (BN-PFO) placed within the cavity, which decreases its refractive index and reduces the cavity optical path length. Significant additional reductions in refractive index and thus blueshift also result from a relatively broad excited state absorption of the BN-PFO that is positioned around 720 nm. We show that the magnitude of the blueshift is directly dependent on the excitation fluence, as is the inverse-lifetime over which this effect persists. Indeed, we demonstrate that a fully-reversible blueshift of the bottom of the LPB of 8 meV can be generated in 750 fs when the cavity is pumped with a fluence of 120 μJ/cm$^2$. The decay of the blueshift is also fast and reaches 50% of its maximum value in under 1 ps. Notably, the blueshift is accompanied by a substantial broadening (factor of 2) of the LPB, which results from the excited state absorption of the out-of-resonance BN-PFO absorber layer. Larger blueshifts of up to 12 meV can be achieved at higher pump fluences; however, these occur at the cost of an irreversible photodegradation of the BN-PFO. Such photo-degradation of the BN-PFO is not, however, expected to affect the BODIPY-Br and would not reduce its ability to undergo polariton condensation and lasing.

The blueshifts observed in this work are significantly larger than those observed in strongly-coupled BODIPY-Br cavities at the condensation threshold which are typically limited to 3-5 meV.[11,38,39] We believe that such large blueshift effects could be used to create energetic barriers in a microcavity that confine or direct the flow of polariton condensates. Such confinement effects could also be harnessed to define dynamic polariton lattices, thereby creating polariton simulator devices or polariton waveguides and routers.[34,35]

**Methods**

*Synthesis of BN-PFO*: BN-PFO, a conjugated polyfluorene-type copolymer with randomly distributed 2,7-(9,9'-di-*n*-octylfluorene) and (2,2'-*n*-octyloxy-1,1'-binaphthyl)-6,6'diyl repeat units (molar content of 6,6'-(2,2'-*n*-octyloxy-1,1'-binaphthyl) repeat units as minority component: 9.8 Mol%; $M_n$: 68,300; $M_w$: 131,000) was synthesized in a nickel(0)-mediated Yamamoto-type homocoupling protocol, starting from the fluorene and binaphthyl monomers as dibromides, by following a procedure reported in [42].

*Thin film fabrication:* BN-PFO was dissolved in toluene at a concentration of 30 mg/mL, with BODIPY-Br dissolved at 10% by mass in a 35 mg/mL solution of polystyrene (PS, Sigma-Aldrich, molecular weight ~192,000) in toluene. Each solution was spin-coated onto a quartz-coated glass substrate to form a thin film. The absorption of each film was measured using a Horiba Fluoromax 4 fluorometer equipped with a xenon lamp. The thickness of the various films was determined using a Bruker DektakXT profilometer. Multilayer films were fabricated by spin-coating the BODIPY-Br/PS solution onto a quartz-coated glass substrate, followed by a 15 mg/mL solution of polyvinyl alcohol (PVA) in deionised water, followed by the BN-PFO solution.

*Cavity fabrication:* A 10-pair $SiO_2/Nb_2O_5$ DBR was fabricated onto a quartz-coated glass substrate by Helia Photonics Ltd using a mix of ion-assisted electron beam evaporation ($Nb_2O_5$) and thermal evaporation ($SiO_2$). A multilayer organic film was spin coated on top of this mirror as described above, with an 8-pair $SiO_2/TiO_2$ top DBR then evaporated on top of this using an Angstrom electron beam deposition system. Here, the deposition chamber contained two

Telemark 25 cc e-beam sources equipped with high-strength graphite crucibles filled either with SiO$_2$ or TiO$_2$, with layers deposited at a rate of 2 Ås$^{-1}$ and a base pressure of 4x10$^{-6}$ mbar.

*Steady-state optical characterisation:* Angle-dependent white light reflectivity measurements were made on cavities using an Ocean Optics Deuterium-Tungsten 378 lamp (DH-2000-BAL) fibre-coupled to the excitation arm of a motorised goniometer setup. The reflected light was collected by an optical fibre on a collection arm, which was connected to an Andor Shamrock SR-303i-A triple-grating CCD spectrometer. The same setup was used to measure the PL spectra of the thin films. In this case, a 405 nm Thorlabs diode laser was focused onto the sample via a third arm that was out-of-plane with respect to the excitation and collection arms

*Transient absorption and reflectivity measurements:* A regeneratively-amplified Ti:sapphire laser emitting 100 fs pulses at 800 nm with a repetition rate of 2 kHz was used to generate the pump and probe beams in both the transmission and reflection configuration. A broadband white light probe beam was generated by focusing the fundamental (800 nm) onto a sapphire crystal. A 400 nm pump beam was produced by second harmonic generation of the 800 nm beam in a β-barium borate crystal, which was then directed through a delay line and a mechanical chopper that reduced the repetition rate to 1 kHz. This was to allow both pump-probe and probe-only measurements. The diameter of the pump beam (435 μm) was chosen such that it was much larger than the probe beam (100 μm diameter), and a ThorLabs camera was used to ensure the entire area being probed was excited by the pump beam.

**Contributions**

K.E.M. characterised the organic dyes and thin films, and designed, fabricated and characterised the cavities with the assistance of R.J. and K.G. under the supervision of D.G.L.. K.E.M. and M.G. performed and analysed the transient absorption and reflectivity measurements under the supervision of T.V. and G.C.. T.J. synthesised the BN-PFO under the supervision of U.S.. A.Z., P.G.L. and D.G.L. conceived the project. All authors contributed to the preparation of the manuscript.

**Acknowledgments**

We thank the U.K. EPSRC for funding this research via the Programme Grant 'Hybrid Polaritonics' (EP/M025330/1). K.E.M. thanks the EPSRC for the award of a Doctoral Training Account PhD studentship (EP/R513313/1) and the Rank Prize Fund for funding a research trip to Politecnico di Milano via a 'Return to Research' grant award. T.V., M.G. and K.E.M. thank

Dr Lucia Ganzer for the useful discussions.

**Conflicts of interest**

**References**

[1]   H. Deng, G. Weihs, D. Snoke, J. Bloch, Y. Yamamoto, *Proceedings of the National Academy of Sciences* **2003**, *100*, 15318.
[2]   S. Christopoulos, G. H. von Högersthal, A. J. D. Grundy, P. G. Lagoudakis, A. V. Kavokin, J. J. Baumberg, G. Christmann, R. Butté, E. Feltin, J. F. Carlin, N. Grandjean, *Phys Rev Lett* **2007**, *98*, 126405.
[3]   G. Christmann, R. Butt{\'e}, E. Feltin, J. F. Carlin, N. Grandjean, *Appl Phys Lett* **2008**, *93*, 1.
[4]   A. Amo, J. Lefrère, S. Pigeon, C. Adrados, C. Ciuti, I. Carusotto, R. Houdré, E. Giacobino, A. Bramati, *Nat Phys* **2009**, *5*, 805.
[5]   E. Wertz, L. Ferrier, D. D. Solnyshkov, P. Senellart, D. Bajoni, A. Miard, A. Lemaître, G. Malpuech, J. Bloch, *Appl Phys Lett* **2009**, *95*, 051108.
[6]   J. Kasprzak, M. Richard, S. Kundermann, A. Baas, P. Jeambrun, J. M. J. Keeling, F. M. Marchetti, M. H. Szymańska, R. André, J. L. Staehli, V. Savona, P. B. Littlewood, B. Deveaud, L. S. Dang, *Nature* **2006**, *443*, 409.
[7]   K. G. Lagoudakis, M. Wouters, M. Richard, A. Baas, I. Carusotto, R. André, L. S. Dang, B. Deveaud-Plédran, *Nat Phys* **2008**, *4*, 706.
[8]   M. Slootsky, Y. Zhang, S. R. Forrest, *Phys Rev B Condens Matter Mater Phys* **2012**, *86*, 1.
[9]   J. D. Plumhof, T. Stöferle, L. Mai, U. Scherf, R. F. Mahrt, *Nat Mater* **2014**, *13*, 247.
[10]  K. S. Daskalakis, S. A. Maier, R. Murray, S. Kéna-Cohen, *Nat Mater* **2014**, *13*, 271.
[11]  T. Cookson, K. Georgiou, A. Zasedatelev, R. T. Grant, T. Virgili, M. Cavazzini, F. Galeotti, C. Clark, N. G. Berloff, D. G. Lidzey, P. G. Lagoudakis, *Adv Opt Mater* **2017**, *5*, 1700203.
[12]  D. Sannikov, T. Yagafarov, K. Georgiou, A. Zasedatelev, A. Baranikov, L. Gai, Z. Shen, D. Lidzey, P. Lagoudakis, *Adv Opt Mater* **2019**, *7*, 1.
[13]  S. Betzold, M. Dusel, O. Kyriienko, C. P. Dietrich, S. Klembt, J. Ohmer, U. Fischer, I. A. Shelykh, C. Schneider, S. Höfling, *ACS Photonics* **2020**, *7*, 384.
[14]  D. Ballarini, M. De Giorgi, E. Cancellieri, R. Houdré, E. Giacobino, R. Cingolani, A. Bramati, G. Gigli, D. Sanvitto, *Nat Commun* **2013**, *4*, 1778.
[15]  H. S. Nguyen, D. Vishnevsky, C. Sturm, D. Tanese, D. Solnyshkov, E. Galopin, A. Lemaître, I. Sagnes, A. Amo, G. Malpuech, J. Bloch, *Phys Rev Lett* **2013**, *110*, 236601.
[16]  F. Marsault, H. S. Nguyen, D. Tanese, A. Lemaître, E. Galopin, I. Sagnes, A. Amo, J. Bloch, *Appl Phys Lett* **2015**, *107*, 201115.
[17]  A. Dreismann, H. Ohadi, Y. Del Valle-Inclan Redondo, R. Balili, Y. G. Rubo, S. I. Tsintzos, G. Deligeorgis, Z. Hatzopoulos, P. G. Savvidis, J. J. Baumberg, *Nat Mater* **2016**, *15*, 1074.
[18]  A. V. Zasedatelev, A. V. Baranikov, D. Urbonas, F. Scafirimuto, U. Scherf, T. Stöferle, R. F. Mahrt, P. G. Lagoudakis, *Nat Photonics* **2019**, *13*, 378.
[19]  F. Chen, H. Li, H. Zhou, S. Luo, Z. Sun, Z. Ye, F. Sun, J. Wang, Y. Zheng, X. Chen, H. Xu, H. Xu, T. Byrnes, Z. Chen, J. Wu, *Phys Rev Lett* **2022**, *129*, 057402.
[20]  H. Deng, H. Haug, Y. Yamamoto, *Rev Mod Phys* **2010**, *82*, 1489.
[21]  A. S. Kuznetsov, K. Biermann, P. V. Santos, *Phys Rev Res* **2019**, *1*, 1.
[22]  G. Lerario, D. Ballarini, A. Fieramosca, A. Cannavale, A. Genco, F. Mangione, S. Gambino, L. Dominici, M. De Giorgi, G. Gigli, D. Sanvitto, *Light Sci Appl* **2016**, *6*, e16212.
[23]  A. Kavokin, T. C. H. Liew, C. Schneider, P. G. Lagoudakis, S. Klembt, S. Hoefling, *Nature*

*Reviews Physics* **2022**, *4*, 435.
- [24] A. Amo, J. Bloch, *C R Phys* **2016**, *17*, 934.
- [25] M. Dusel, S. Betzold, O. A. Egorov, S. Klembt, J. Ohmer, U. Fischer, S. Höfling, C. Schneider, *Nat Commun* **2020**, *11*, 1.
- [26] F. Scafirimuto, D. Urbonas, M. A. Becker, U. Scherf, R. F. Mahrt, T. Stöferle, *Commun Phys* **2021**, *4*, 39.
- [27] M. Obert, J. Renner, A. Forchel, G. Bacher, R. André, D. Le Si Dang, *Appl Phys Lett* **2004**, *84*, 1435.
- [28] A. M. Adawi, A. Cadby, L. G. Connolly, W. C. Hung, R. Dean, A. Tahraoui, A. M. Fox, A. G. Cullis, D. Sanvitto, M. S. Skolnick, D. G. Lidzey, *Advanced Materials* **2006**, *18*, 742.
- [29] D. Bajoni, P. Senellart, E. Wertz, I. Sagnes, A. Miard, A. Lemaître, J. Bloch, *Phys Rev Lett* **2008**, *100*, 1.
- [30] R. Jayaprakash, C. E. Whittaker, K. Georgiou, O. S. Game, K. E. McGhee, D. M. Coles, D. G. Lidzey, *ACS Photonics* **2020**, *7*, 2273.
- [31] M. Wei, W. Verstraelen, K. Orfanakis, A. Ruseckas, T. C. H. Liew, I. D. W. Samuel, G. A. Turnbull, H. Ohadi, *Nat Commun* **2022**, *13*, 7191.
- [32] R. Balili, V. Hartwell, D. Snoke, L. Pfeiffer, K. West, *Science (1979)* **2007**, *316*, 1007.
- [33] R. Balili, B. Nelsen, D. W. Snoke, L. Pfeiffer, K. West, *Phys Rev B Condens Matter Mater Phys* **2009**, *79*, 1.
- [34] G. Tosi, G. Christmann, N. G. Berloff, P. Tsotsis, T. Gao, Z. Hatzopoulos, P. G. Savvidis, J. J. Baumberg, *Nat Phys* **2012**, *8*, 190.
- [35] P. Cristofolini, A. Dreismann, G. Christmann, G. Franchetti, N. G. Berloff, P. Tsotsis, Z. Hatzopoulos, P. G. Savvidis, J. J. Baumberg, *Phys Rev Lett* **2013**, *110*, 186403.
- [36] T. Yagafarov, D. Sannikov, A. Zasedatelev, K. Georgiou, A. Baranikov, O. Kyriienko, I. Shelykh, L. Gai, Z. Shen, D. Lidzey, P. Lagoudakis, *Commun Phys* **2020**, *3*, 1.
- [37] Y. Kadoya, K. Kameda, M. Yamanishi, T. Nishikawa, T. Kannari, T. Ishihara, I. Ogura, *Appl Phys Lett* **1995**, *281*, 281.
- [38] A. Putintsev, A. Zasedatelev, K. E. McGhee, T. Cookson, K. Georgiou, D. Sannikov, D. G. Lidzey, P. G. Lagoudakis, *Appl Phys Lett* **2020**, *117*, 123302.
- [39] K. E. McGhee, A. Putintsev, R. Jayaprakash, K. Georgiou, M. E. O'Kane, R. C. Kilbride, E. J. Cassella, M. Cavazzini, D. A. Sannikov, P. G. Lagoudakis, D. G. Lidzey, *Sci Rep* **2021**, *11*, 20879.
- [40] R. T. Grant, P. Michetti, A. J. Musser, P. Gregoire, T. Virgili, E. Vella, M. Cavazzini, K. Georgiou, F. Galeotti, C. Clark, J. Clark, C. Silva, D. G. Lidzey, *Adv Opt Mater* **2016**, *4*, 1615.
- [41] A. J. Musser, S. K. Rajendran, K. Georgiou, L. Gai, R. T. Grant, Z. Shen, M. Cavazzini, A. Ruseckas, G. A. Turnbull, I. D. W. Samuel, J. Clark, D. G. Lidzey, *J Mater Chem C Mater* **2017**, *5*, 8380.
- [42] T. Rabe, M. Hoping, D. Schneider, E. Becker, H. H. Johannes, W. Kowalsky, T. Weimann, J. Wang, P. Hinze, B. S. Nehls, U. Scherf, T. Farrell, T. Riedl, *Adv Funct Mater* **2005**, *15*, 1188.
- [43] M. Lehnhardt, T. Riedl, U. Scherf, T. Rabe, W. Kowalsky, *Org Electron* **2011**, *12*, 1346.
- [44] T. Virgili, D. G. Lidzey, D. D. C. Bradley, *Advanced Materials* **2000**, *12*, 58.
- [45] V. Savona, L. C. Andreani, P. Schwendimann, A. Quattropani, *Solid State Commun* **1995**, *93*, 733.
- [46] G. Cerullo, S. Stagira, M. Zavelani-Rossi, S. De Silvestri, T. Virgili, D. G. Lidzey, D. D. C. Bradley, *Chem Phys Lett* **2001**, *335*, 27.
- [47] H. Marciniak, M. Teicher, U. Scherf, S. Trost, T. Riedl, M. Lehnhardt, T. Rabe, W. Kowalsky, S. Lochbrunner, *Phys Rev B* **2012**, *85*, 214204.
- [48] T. Virgili, D. Marinotto, C. Manzoni, G. Cerullo, G. Lanzani, *Phys Rev Lett* **2005**, *94*, 117402.
- [49] S. Xu, V. I. Klimov, B. Kraabel, H. Wang, D. W. McBranch, *Phys Rev B* **2001**, *64*, 193201.
- [50] R. P. Sabatini, T. M. McCormick, T. Lazarides, K. C. Wilson, R. Eisenberg, D. W. McCamant, *J Phys Chem Lett* **2011**, *2*, 223.
- [51] X.-F. Zhang, X. Yang, *J Phys Chem B* **2013**, *117*, 5533.

[52] J. Al Anshori, T. Slanina, E. Palao, P. Klán, *Photochemical & Photobiological Sciences* **2016**, *15*, 250.
[53] K. Yamashita, U. Huynh, J. Richter, L. Eyre, F. Deschler, A. Rao, K. Goto, T. Nishimura, T. Yamao, S. Hotta, H. Yanagi, M. Nakayama, R. H. Friend, *ACS Photonics* **2018**, *5*, 2182.
[54] M. S. Skolnick, T. A. Fisher, D. M. Whittaker, *Semicond Sci Technol* **1998**, *13*, 645.